# Abstracting Runtime Heaps for Program Understanding

Mark Marron[1]    Cesar Sanchez[1,2]    Zhendong Su[3]    Manuel Fahndrich[4]

[1]IMDEA Software Institute   [2]CSIC   [3]UC Davis   [4]Microsoft Research
{mark.marron, cesar.sanchez}@imdea.org, su@ucdavis.edu, maf@microsoft.com

**Abstract.** Modern programming environments provide extensive support for inspecting, analyzing, and testing programs based on the algorithmic structure of a program. Unfortunately, support for inspecting and understanding runtime data structures during execution is typically much more limited. This paper provides a general purpose technique for abstracting and summarizing entire runtime heaps. We describe the abstract heap model and the associated algorithms for transforming a concrete heap dump into the corresponding abstract model as well as algorithms for merging, comparing, and computing changes between abstract models. The abstract model is designed to emphasize high-level concepts about heap-based data structures, such as shape and size, as well as relationships between heap structures, such as sharing and connectivity. We demonstrate the utility and computational tractability of the abstract heap model by building a memory profiler. We then use this tool to check for, pinpoint, and correct sources of memory bloat from a suite of programs from DaCapo.

## 1  Introduction

Modern programming environments provide excellent support for visualizing and debugging code, but inspecting and understanding the high-level structure of the data manipulated at runtime by said code is typically not well supported. Visualizing entire runtime heap graphs is a non-trivial problem, as the number of nodes and edges is typically so large that displaying these graphs directly—even with excellent graph visualization tools—results in useless jumbles of nodes and edges. As a result, little of interest can be gleaned from such visualizations.

In this paper, we propose an abstract domain for runtime heap graphs that captures many fundamental properties of data structures on the heap, such as shape, connectivity, and sharing, but abstracts away other often less useful details. The abstract heap graphs we compute are both small enough to visualize and navigate, and at the same time precise enough to capture essential information useful in interactive debugging and memory profiling scenarios. Further, the abstract heaps can be computed efficiently from a single concrete heap and further merged/compared with other abstract heap graphs. from across a set of program runs, or from multiple program points in order to get an even more general view of the heap configurations that occur during program execution.

*Example.* Figure 1(a) shows a heap snapshot of a simple program that manipulates expression trees. An expression tree consists of binary nodes for `Add`, `Sub`, and `Mult`, and leaf nodes for `Constants` and `Variables`. The local variable `exp` (rectangular box) points to an expression tree consisting of 4 interior binary expression objects, 2 `Var`, and 2 `Const` objects. Local variable `env` points to an array representing an environment of `Var` objects that are shared with the expression tree.

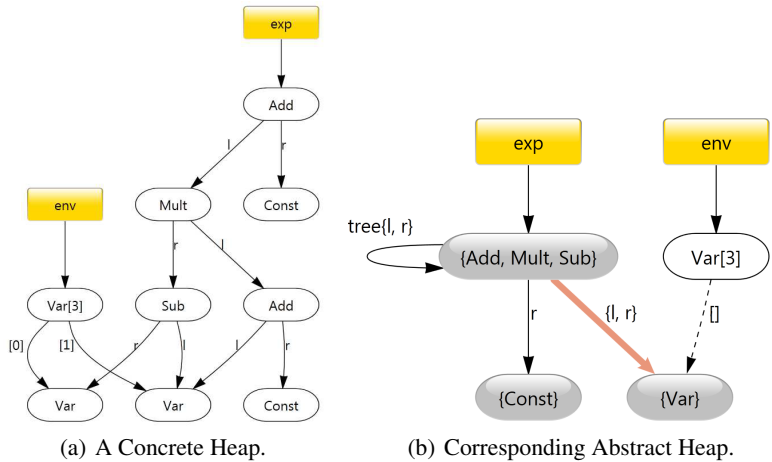

(a) A Concrete Heap.  (b) Corresponding Abstract Heap.

**Fig. 1.** A concrete heap and corresponding abstraction.

Figure 1(b) shows the abstract heap produced by our tools from this concrete heap with the default visualization mode.[1] The abstraction summarizes the concrete objects into three distinct summary nodes in the abstract heap graph: (1) an abstract node representing all interior recursive objects in the expression tree (`Add`, `Mult`, `Sub`), (2) an abstract node representing the two `Var` objects, and (3) an abstract node representing the two `Const` objects. Specific details about the order and branching structure of expression nodes are absent in the abstraction, but other more general properties are still present. For example, the fact that there is no sharing or cycles among the interior expression nodes is apparent in the abstract graph by looking at the self-edge representing the pointers between objects in the interior of the expression tree. The label `tree{l,r}` on the self-edge expresses that pointers stored in the `l` and `r` fields of the objects in this region form a tree structure (i.e., no sharing and no cycles).

The abstract graph maintains another useful property of the expression tree, namely that no `Const` object is referenced from multiple expression objects. On the other hand, several expression objects might point to the same `Var` object. The abstract graph shows possible sharing using wide orange colored edges (if color is available), whereas normal edges indicate non-sharing pointers. The abstract graph shows pointer nullity via full vs. dashed lines – in our example all pointers, except in the environment array, are non-null.

Rudimentary information on the number of objects represented by each node is encoded in the shading. Nodes that always abstract a single object are given a white background while nodes which represent multiple objects are shaded (silver if color is available). Size information of arrays and other containers is encoded by annotating the type label with the container size (`Var[3]` to indicate an array is of length 3).

---

[1] Additional information can be obtained by hovering over the nodes/edges or by restyling for a specific task as in our case studies in section 6.

*Overview.* This paper addresses the problem of turning large concrete runtime heaps into compact abstract heaps while retaining many interesting properties of the original heap in the abstraction. Our abstraction is safe in the sense that properties stated on the abstract heap graph also hold in the corresponding concrete heaps. To achieve this abstraction safety, we adopt the theory for the design of abstract domains developed in *abstract interpretation* [7,25]. The theory of abstract interpretation provides a general framework for (1) defining an abstract domain and relating it to possible concrete program states and (2) a method for taking an abstract domain and computing an over-approximation of the collecting semantics for a given program as a static analysis. The static analysis component of the abstract interpretation framework is not relevant here, as we are interested in abstracting runtime heaps. However, the framework for constructing the abstract domains, as well as the properties of operations for comparing ($\sqsubseteq$) and merging ($\widetilde{\sqcup}$) abstract domain elements, allows us to formally describe the relationship of our abstract heap graphs to their concrete counterparts, and to obtain safe operations for comparing and summarizing heaps from different program points or different program runs in a semantically meaningful way. These guarantees provide confidence that all inferences made by examining the abstract model are valid.

Our abstract heap domain encodes a fixed set of heap properties identified in previous work on static heap analysis [5,10,20] that are fundamental properties of heaps and can be computed efficiently. These properties include the summarization of recursive and composite data structures, the assignment of shape information to these structures and injectivity of fields (given two distinct objects does the field `f` in each object point to a distinct target). The abstraction is also able to provide information on the number and types of objects in the various structures, as well as nullity information. Our focus on a fixed set of heap properties (as opposed to user defined properties) enables the abstraction to be computed efficiently in time $O((Ob+Pt)*log(Ob))$, where $Ob$ is the number of objects and $Pt$ is the number of pointers in the concrete heap.

The contributions of this paper are:

- The abstract domain for heap graphs and its concretization function formalizing the safe relationship to concrete heaps.
- An efficient algorithm for computing the abstraction and algorithms for comparing and joining abstract heaps.
- Graphical representations of abstract heap graphs that allow on-demand collapsing or expansion of sub-structures, allowing a form of *semantic zoom* [8] from a very abstract view of the heap down to the level of individual objects.
- The construction of a general purpose heap memory profiler and analysis tool that augments the basic abstraction with specialized support for profiling and identifying common memory problems in a program.
- A qualitative evaluation of the visualization and memory profiler in tracking down and identifying solutions to memory inefficiencies in a range of programs (up to a 25% reduction in memory use).

## 2 Abstract Heap Graph

We begin by formalizing concrete program heaps and the relevant properties of concrete heaps that will be captured by the abstraction. Later, we define the abstract heap graph and formally relate the abstraction to its concrete heap counterparts using a *concretization* ($\gamma$) function from the framework of abstract interpretation.

### 2.1 Concrete Heaps

For the purposes of this paper, we model the runtime state of a program as an environment, mapping variables to values, and a store, mapping addresses to values. We refer to an instance of an environment together with a store as a *concrete heap*. Formally, a concrete heap is a labeled directed graph $(\text{root}, \text{null}, Ob, Pt, Ty)$, where the nodes are formed by the set of heap objects ($Ob$) and the edges ($Pt$) correspond to pointers. We assume a distinguished heap object $\text{root} \in Ob$ whose fields are the variables from the environment. This representation avoids dealing with distinct sets of variable locations and makes the formalization more uniform. We also assume a distinguished object $\text{null}$ among $Ob$ to model null pointers. The set of pointers $Pt \subseteq Ob \times Ob \times \text{Label}$ connect a source object to a target object with a pointer label from $\text{Label}$. These labels are either a variable name (if the source object is $\text{root}$), a field name (if the source object is a heap object), or an array index (if the source object is an array). Finally, $Ty : Ob \to \text{Type}$ is a map that assigns a concrete program type to each object. We assume the concrete set of types in $\text{Type}$ contains at least object types and array types. We use the notation $o_1 \xrightarrow{p} o_2$ to indicate that object $o_1$ refers to $o_2$ via pointer label $p$.

A *region* of memory $C \subseteq Ob \setminus \{\text{null}, \text{root}\}$ is a subset of the concrete heap objects, not containing the root node or null. It is handy to define the set of pointers $P(C_1, C_2)$ crossing from a region $C_1$ to a region $C_2$ as:

$$P(C_1, C_2) = \{o_1 \xrightarrow{p} o_2 \in Pt \mid o_1 \in C_1, o_2 \in C_2\}$$

### 2.2 Concrete Heap Properties

We now formalize the set of concrete properties of objects, pointers, and entire regions of the heap that we later use to create the abstract heap graph.

*Type.* The set of types associated with a region $C$ is the union of all types of the objects in the region: $\{Ty(o) \mid o \in C\}$.

*Cardinality.* The cardinality of a region $C$ is the number of objects in the region $|C|$.

*Nullity.* A pointer $o_1 \to o_2$ is a null pointer if $o_2 = \text{null}$ and non-null pointer if $o_2 \neq \text{null}$.

*Injectivity.* Given two regions $C_1$ and $C_2$, we say that pointers labeled $p$ from $C_1$ to $C_2$ are *injective*, written $\text{inj}(C_1, C_2, p)$, if for all pairs of pointers $o_1 \xrightarrow{p} t_1$ and $o_2 \xrightarrow{p} t_2$ drawn from $P(C_1, C_2)$, $o_1 \neq o_2 \Rightarrow t_1 \neq t_2$. In words, the pointers labeled $p$ from two *distinct* objects $o_1$ and $o_2$ point to *distinct* objects $t_1$ and $t_2$.

*Shape.* We characterize regions of memory *C* by shape using standard graph theoretic notions of trees and general graphs. For additional precision, we consider the shape of subgraphs formed from *C*, and $P(C,C)\!\downarrow_L$, i.e., the subgraph consisting of objects from *C* and pointers with labels $l \in L$ only. This way, we can describe, for example, that a tree structure with parent pointers is still a tree if we only consider the left and right pointers, but not the parent pointers.

- The predicate $\mathsf{any}(C,L)$ is simply true for any graph. We use it only to clarify shapes in visualizations that don't satisfy the more restrictive tree property.
- The predicate $\mathsf{tree}(C,L)$ holds, if $P(C,C)\!\downarrow_L$ is acyclic and the subgraph $P(C,C)\!\downarrow_L$ does not contain any cross edges.

### 2.3 Heap Graph Abstraction

An abstract heap graph is an instance of storage shape graphs [5]. More precisely, the abstract heap graphs used in this paper are tuples:

$$(\mathsf{root}, \mathsf{null}, Ob^\#, Pt^\#, Ty^\#, Cd^\#, Ij^\#, Sh^\#)$$

where $Ob^\#$ is a set of abstract nodes (each of which abstracts a region of the concrete heap), and $Pt^\# \subseteq Ob^\# \times Ob^\# \times \mathsf{Label}^\#$ is a set of graph edges, each of which abstracts a set of pointers. Edges are annotated with labels from, $\mathsf{Label}^\#$, and are the field labels and the special label $[]$. The special label $[]$ abstracts the indices of all array or container elements (i.e., array smashing).

We distinguish a root node in $Ob^\#$ for modeling the variable environment as fields on root. Another distinguished node null is used to represent the null pointer. The remaining parts of an abstract heap $(Ty^\#, Cd^\#, Ij^\#, Sh^\#)$ capture abstract properties of the heap graph. $Ty^\# : Ob^\# \mapsto 2^{\mathsf{Type}}$ maps abstract nodes to the set of types of the concrete nodes represented by the abstraction. $Cd^\# : Ob^\# \mapsto \mathit{Interval}$ represents the cardinality of each abstracted region. $Cd^\#$ maps each abstract node *n* to a numerical interval $[l,u] \in \mathit{Interval}$, where the lowerbound *l* is a natural number, and *u* is a natural number or $\infty$.

The abstract injectivity $Ij^\# : Pt^\# \to \mathsf{bool}$ expresses whether the set of pointers represented by an abstract edge is injective. Finally, the abstract shape $Sh^\#$ is a set of tuples $(n, L, s) \in Ob^\# \times 2^{\mathsf{Label}^\#} \times \{\mathsf{tree}, \mathsf{any}\}$ indicating the shape *s* of a region represented by *n* with edges restricted to *L*.

### 2.4 Abstraction Relation

We are now ready to formally relate the abstract heap graph to its concrete counterparts by specifying which heaps are in the concretization of an abstract heap:

$$\begin{aligned}
(\mathsf{root}, \mathsf{null}, Ob, Pt, Ty) \in {}& \gamma(\mathsf{root}, \mathsf{null}, Ob^\#, Pt^\#, Ty^\#, Cd^\#, Ij^\#, Sh^\#) \Leftrightarrow \\
& \exists \mu \,.\, \mathsf{Embed}(\mu, Ob, Pt, Ob^\#, Pt^\#) \\
& \wedge \mathsf{Typing}(\mu, Ob, Ty, Ob^\#, Ty^\#) \wedge \mathsf{Counting}(\mu, Ob, Ob^\#, Cd^\#) \\
& \wedge \mathsf{Injective}(\mu, Pt, Pt^\#, Ij^\#) \wedge \mathsf{Shape}(\mu, Pt, Pt^\#, Sh^\#)
\end{aligned}$$

A concrete heap is an instance of an abstract heap, if there exists an embedding $\mu : Ob \to Ob^{\#}$ satisfying the graph embedding, typing, counting, injectivity, and shape relation between the graphs. The auxiliary predicates are defined as follows.

$$\mathsf{Embed}(\mu, Ob, Pt, Ob^{\#}, Pt^{\#}) \Leftrightarrow \mu(\mathsf{root}) = \mathsf{root} \wedge \mu(\mathsf{null}) = \mathsf{null}$$

$$\wedge \forall o_1 \xrightarrow{p} o_2 \in Pt . \exists l . \mu(o_1) \xrightarrow{l} \mu(o_2) \in Pt^{\#} \wedge p \in \gamma_L(l)$$

The embed predicate makes sure that all edges of the concrete graph are present in the abstract graph, connecting corresponding abstract nodes, and that the edge label in the abstract graph encompasses the concrete edge label. The embedding mapping $\mu$ must also map the special objects root and null to their exact abstract counterparts.

$$\mathsf{Typing}(\mu, Ob, Ty, Ob^{\#}, Ty^{\#}) \Leftrightarrow \forall o \in Ob . Ty(o) \in Ty^{\#}(\mu(o))$$

The typing relation guarantees that the type $Ty(o)$ for every concrete object $o$ is in the set of types $Ty^{\#}(\mu(o))$ of the abstract node $\mu(o)$ of $o$.

$$\mathsf{Counting}(\mu, Ob, Ob^{\#}, Cd^{\#}) \Leftrightarrow \forall n \in Ob^{\#} . |\mu^{-1}(n)| \in Cd^{\#}(n)$$

The counting relation guarantees that for each abstract node $n$, the set of concrete nodes $\mu^{-1}(n)$ abstracted by $n$ has a cardinality in the numeric interval $Cd^{\#}(n)$.

$$\mathsf{Injective}(\mu, Pt, Pt^{\#}, Ij^{\#}) \Leftrightarrow$$

$$\forall (n_1, n_2, l) \in Pt^{\#} . Ij^{\#}(n_1, n_2, l) \Rightarrow \forall p \in \gamma_L(l) . \mathsf{inj}(\mu^{-1}(n_1), \mu^{-1}(n_2), p)$$

The injectivity relation guarantees that every pointer set marked as injective corresponds to injective pointers between the concrete source and target regions of the heap.

$$\mathsf{Shape}(\mu, Pt, Pt^{\#}, Sh^{\#}) \Leftrightarrow \forall (n, L, \mathsf{tree}) \in Sh^{\#} . \mathsf{tree}(\mu^{-1}(n), \gamma_L(L))$$

Finally, the shape relation guarantees that for every abstract shape tuple $(n, L, s)$, the concrete subgraph $\mu^{-1}(n)$ abstracted by node $n$ restricted to labels $L$ satisfies the corresponding concrete shape predicate $s$ (tree and implicitly any).

### 2.5 Visual Representation of Abstract Heap Graphs

In the iconography for our abstract graph visualizations, the screen shots in Figure 1(a), Figure 1(b), and section 6, we leverage a number of conventions to convey information.

An edge $(\mathsf{root}, o, p)$ whose source is the root node represents the content of variable $p$. Instead of drawing a root node with such edges, we simply draw a variable node $p$ and an unlabeled edge to $o$. Thus, the root node is never drawn, as it does not appear as the target of any edge in concrete or abstract graphs.

The set of abstract types of an abstract node is represented as the label of the abstract node. Shape information is represented as labels on the recursive self edges of abstract nodes. An abstract node with cardinality 1 is represented by a white background. Other cardinalities are represented with shaded abstract nodes.

We do not draw explicit edges which only point to null. If an edge is associated with a label that contains both pointers to null and pointers to other heap objects we fold the possibility into the edge by using a dashed edge instead of a full edge. Finally, injective edges are represented with normal thin edges, whereas non-injective edges are represented by wide edges (and if color is available are also highlighted in orange).

## 3 Computing the Abstraction

This section describes the computation of the abstract graph from a given concrete heap. The transformation is performed in three phases. 1) recursive data structures are identified and collapsed based on identifying cycles in the type definitions, 2) nodes that represent objects in the same logical heap region based on equivalent edges originating the same abstract node are merged, and finally 3) abstract properties like cardinality, injectivity, and shape are computed for the abstract edges and nodes.

### 3.1 Partition ($\mu$) Computation

Initially, we associate with each concrete object $o_i$ an abstract partition $n_i$ representing an equivalence using a Tarjan union-find structure. The mapping $\mu$ from concrete objects to abstract partitions is given at any point in time by: $\mu(o_i) = \text{ecr}(n_i)$, i.e., by the equivalence class of the original $n_i$ associated with $o_i$. The union-find structure maintains the reverse mapping $\mu^{-1}$ providing the set of concrete objects abstracted by a node. The abstract type map $Ty^{\#}$ can be maintained efficiently in the union-find structure as well.

Figure 2(a) shows the initial state of these equivalence partitions for our example from Figure 1(a) (one partition per object, plus the roots, and a special partition for null). Each node is labeled with its partition id and the types of the objects in that partition.

The first abstraction identifies parts of the heap graph that represent unbounded depth recursive data structures. The basic approach consists of examining the type information in the program and the heap connectivity properties [2,19,9] and ensures that any heap graph produced has a finite depth. We say types $\tau_1$ and $\tau_2$ are *recursive* ($\tau_1 \sim \tau_2$) if they are part of the same recursive type definition.

**Definition 1 (Same Data Structure Objects).** *Two distinct objects $o_1$, $o_2$ are part of the same data structure if there is a reference $o_1 \xrightarrow{p} o_2$ in the heap and the types of the two objects are in the same data structure $Ty(o_1) \sim Ty(o_2)$.*

The recursive components are thus identified by visiting each pointer $o_i \to o_j$ in the heap and if $o_i$ and $o_j$ are in the same data structure according to Definition 1, then we union the corresponding abstract nodes $n_i$ and $n_j$.

Figure 2(b) shows the result of merging *Same Data Structure Nodes* on the initial partitions shown in Figure 2(a). The algorithm identified objects $1, 2, 4, 5$ (the `Add`, `Sub`, and `Mult` objects from the interior of the expression tree) as being part of the recursive data structure and replaced them with a single representative summary node.

Next we group objects based on predecessor partitions. The motivation for this abstraction can be seen in Figure 2(b) where `Var` objects in partitions 7 and 8 represent "variables in the environment". There's no need to distinguish them as they are both referenced from the environment array. Similarly, the two constant objects referenced from the recursive component both represent "constants in the expression tree".

**Definition 2 (Equivalent on Abstract Predecessors).** *Given two pointers $o_1 \xrightarrow{l} o_2$ and $o'_1 \xrightarrow{l'} o'_2$ where $\mu(o_1) = \mu(o'_1)$, we say that their target nodes are equivalent whenever: The labels agree $l = l'$, and the target nodes have some types in common, i.e., $Ty^{\#}(\mu(o_2)) \cap Ty^{\#}(\mu(o'_3)) \neq \emptyset$.*

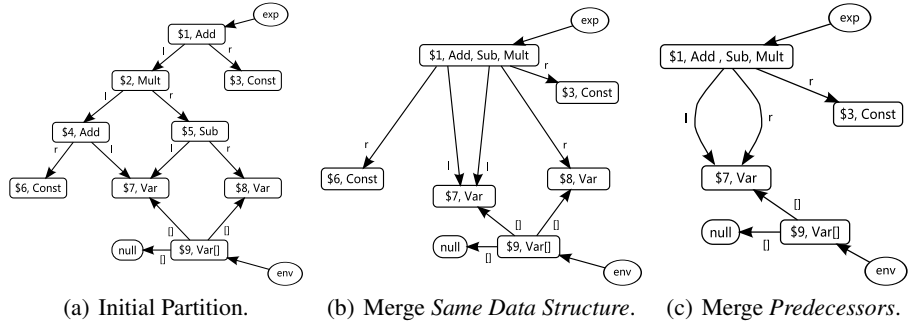

(a) Initial Partition.  (b) Merge *Same Data Structure*.  (c) Merge *Predecessors*.

**Fig. 2.** Steps in abstraction computation.

The algorithm for grouping equivalent objects is based on a worklist where merging two partitions may create new opportunities for merging. The worklist consists of pointers that may have *equivalent* target objects. When processing a pointer from the worklist we check if we need merge any partitions and, as needed, we merge these partitions. Finally, all pointers incident to the merged partitions are added to the worklist. Due to the properties of the Tarjan union-find algorithm, each pointer can enter the work list at most $log(N)$ times, where $N$ is the number of abstract partitions that can be merged, and $E$ is the number of pointers. Thus the complexity of this step is $O(E * log(N))$.

Figure 2(c) shows the result of performing the required merge operations on the partitions from Figure 2(b). The algorithm has merged the Var regions into a new summary region (since the objects represented by partitions 7 and 8 in Figure 2(b) are referred to from the same array). Similarly the Const partitions from Figure 2(b) have been merged as they are both stored in the same recursive structure (the expression tree).

Figure 2(c) differs from the abstract graph in Figure 1(b) where there is only one edge between the expression tree node and the variables node. The reason is that, despite the underlying abstraction being a multi-graph, our visualization application collapses multi-edges as they frequently lead to poor graph layouts and rarely provide useful information to the developer. Also, note that there are explicit references to null and that these were not merged since we associate no types with the null object.

### 3.2 Abstract Property Computation

*Type, Cardinality, and Nullity.* The abstract type map $Ty^{\#}$ has already been computed as part of the union-find operation on abstract nodes. Similarly, the union-find operation computes the exact cardinality, which results in a precise interval value $[i,i]$ if a node abstracts exactly $i$ objects. The nullity information is represented as explicit edges to the null abstract object.

*Injectivity.* The *Injectivity* information for an abstract edge $n_1 \xrightarrow{l} n_2$ is computed by iterating over all pointers from objects $o_i$ represented by $n_1$ to objects $o_j$ represented by $n_2$ with label $p$ compatible with $l$. We determine if every concrete target object is referenced at most once, in which case the abstract edge is *injective*. Otherwise, the edge is *not injective*.

*Shape.* The fundamental observation that enables interesting shape predicates to be produced for the abstract graphs is that the shape properties are restricted to the subgraphs represented by an abstract node. In addition, we allow the examination of a variety of further subgraphs by restricting the set of labels considered in the subgraph. Restricting the label set allows e.g., to determine that the {l,r} edges in a tree actually form a tree, even though there are also parent pointers p, which if included would allow no interesting shape property to be determined. Selecting the particular subsets of edge labels to consider in the subgraph selection is based on heuristics. We can start with all labels to get an overall shape and use that computation to guess which labels to throw out and try again. For small sets of labels, all combinations can be tried.

After partitioning the heap as shown in Figure 2(c) the final map for the objects is:

$$\mu^{-1} = \{\, n_1 \mapsto \{o_1,o_2,o_4,o_5\}, n_3 \mapsto \{o_3,o_6\}, n_7 \mapsto \{o_7,o_8\}, n_9 \mapsto \{o_9\}\,\}$$

Thus, for Figure 1(b) we determine the abstract edge representing the cross partition pointer set $n_1 \xrightarrow{l} n_7$ is *not injective*, since it abstracts the two concrete pointers $o_4 \xrightarrow{l} o_7$ and $o_5 \xrightarrow{l} o_7$ both refer to the same Var object $o_7$. On the other hand, since the two Const objects $o_3$, $o_6$ are distinct, the algorithm will determine that edge representing the cross partition pointer set $n_1 \xrightarrow{r} n_3$ is *injective*. The *Shape* computation for the node representing partition 1 requires a traversal of the four objects. As there are no cross or back edges the layout for this is tree{l,r}.

## 4 Merge and Comparison Operations

Many program analysis and understanding tasks require the ability to 1) accumulate abstract graphs, and 2) compare abstract graphs (both from the same program execution and across executions). For example, to support computing differences in the heap state during profiling activities or for computing likely heap invariants. So we cannot simply track object identities and use them to control the merge and compare operations. Thus, the definitions must be entirely based on the abstract graph structure.

### 4.1 Compare

Formally, the order between two abstract graphs $g_1 \sqsubseteq g_2$ can be defined via our abstraction relation from subsection 2.4 as: $g_1 \sqsubseteq g_2 \Leftrightarrow \forall h.h \in \gamma(g_1) \Rightarrow h \in \gamma(g_2)$.

However, this is not directly computable. Instead, we implement a $O(E)$ time approximation of this relation that first determines the structural equality of the abstract graphs by computing an isomorphism, followed by an implication check that all abstract edge and node properties in $g_2$ cover the equivalent node and edge properties of $g_1$.

To efficiently compute the subgraph isomorphism between $g_1$ and $g_2$ we use a property of the abstract graphs established by Definition 2. From this definition we know that every pair of out edges from a node either differ in the *label* or have the same *label* but non-overlapping sets of *types* in the nodes they refer to. Thus, to compute an isomorphism between two graphs we can simply start pairing the local and global roots and then from each pair match up edges based on their *label* and *type* sets, leading to new pairings. This either results in an isomorphism map, or it results in a pair of nodes reachable from the roots along the same path that have incompatible edges. Any such edge differences can then be reported. With the subgraph isomorphism $\phi$, we define the ordering relation:

$$g_1 \sqsubseteq_\phi g_2 \Leftrightarrow \forall n \in Ob_1^\#.Ty_1^\#(n) \subseteq Ty_2^\#(\phi(n))$$
$$\land \forall n \in Ob_1^\#.Cd_1^\#(n) \sqsubseteq Cd_2^\#(\phi(n)) \land \forall \phi(e) \in Pt_2^\#.Ij_2^\#(\phi(e)) \Rightarrow Ij_1^\#(e)$$
$$\land \forall (\phi(n), L_2, s_2) \in Sh_2^\#.\exists (n, L_1, s_1) \in Sh_1^\#.L_2 \subseteq L_1 \land s_1 \sqsubseteq s_2$$

Note how abstract shape predicates are contra-variant in the label set *L*. In other words, if a shape property holds for the subgraph based on $L_1$, then it holds for the smaller subgraph based on the smaller set $L_2$.

### 4.2 Merge

The merge operation takes two abstract graphs and produces a new abstract graph that is an over approximation of all the concrete heap states that are represented by the two input graphs. In the standard abstract interpretation formulation this is typically the least element that is an over approximation of both models. However, to simplify the computation we do not enforce this property (formally we define an *upper approximation* instead of a *join*). Our approach is to leverage the existing definitions from the abstraction function in the following steps.

Given two abstract heap graphs, $g_1$ and $g_2$ of the form $g_i = (\text{root}_i, \text{null}_i, Ob_i^\#, Pt_i^\#, Ty_i^\#, Cd_i^\#, Ij_i^\#, Sh_i^\#)$ we can define the graph, $g_3$, that is the result of their merge as follows. First we produce the union of the two graphs by simply adding all nodes and edges from both graphs. Once we have taken the union of the two graphs we merge the variable/static roots that have the same names. Then we use Definition 1 and Definition 2 to zip down the graph merging nodes and edges until no more changes are occurring. During the merge we build up two mappings $\eta_1 : g_1 \to g_3$ and $\eta_2 : g_2 \to g_3$ from nodes (edges) in the original graphs, $g_1$ and $g_2$ respectively, to the nodes (edges) in the merged graph. Using these mappings, we define upper approximations of all the graph properties:

$$Ty_3^\#(n) = \bigcup_{n_1 \in \eta_1^{-1}(n)} Ty_1^\#(n_1) \cup \bigcup_{n_2 \in \eta_2^{-1}(n)} Ty_2^\#(n_2)$$

$$Cd_3^\#(n) = \sum_{n_1 \in \eta_1^{-1}(n)} Cd_1^\#(n_1) \sqcup \sum_{n_2 \in \eta_2^{-1}(n)} Cd_2^\#(n_2)$$

$$Ij_3^\#(e) = (|\eta_1^{-1}(e)| = |\{n_2 \mid n_1 \xrightarrow{l} n_2 \in \eta_1^{-1}(e)\}|)$$
$$\land (|\eta_2^{-1}(e)| = |\{n_2 \mid n_1 \xrightarrow{l} n_2 \in \eta_2^{-1}(e)\}|)$$
$$\land \bigwedge_{e_1 \in \eta_1^{-1}(e)} Ij_1^\#(e_1) \land \bigwedge_{e_2 \in \eta_2^{-1}(e)} Ij_2^\#(e_2)$$

The set of types associated with the result is just the union of all types abstracted by the node in both graphs. The cardinality is more complicated to compute. It computes the abstract sums over intervals from all nodes abstracted from the input graphs separately, and then joins the resulting interval (or depending on the application widens as defined in [7]). Injectivity is the logical conjunction of the injectivity of all the source edges, provided that all the edges in the respective graphs that are merged had different target nodes (the equality of the edge and target sets). When merging two injective edges from the same graph we cannot guarantee that the resulting set of edges is injective, in the case that they target the same node, and if we encounter this we conservatively assume the result edge is *not injective*.

For computing the *shape* predicates we need to take into account not only the shape properties of the original graphs, but also the connectivity among the input nodes that map to the same node in the joined graph. We define a very conservative check for treeness during the merge:

$$\mathsf{tree}_\mu(n, L, \mu, g) \Leftrightarrow |Pt^{\#}_g\!\downarrow_{\mu^{-1}(n), L}| \leq 1 \wedge \forall n' \in \mu^{-1}(n). \exists L' \supseteq L(n', L', \mathsf{tree}) \in Sh^{\#}_g$$

where $Pt^{\#}_g\!\downarrow_{\mu_i^{-1}(n), L}$ is the subgraph of $Pt^{\#}_g$ made up of nodes that map to $n$ under $\mu$ and non-self[2] edges incident to them and restricted to labels $L$. Note that tree can only be inferred, if at most one node is in the partition from each graph and the node represents a tree. The abstract shape for a merged node in the graph can be defined as:

$$(n, L, \mathsf{tree}) \in Sh^{\#}_3 \Leftrightarrow \mathsf{tree}_\mu(n, L, \mu, g_1) \wedge \mathsf{tree}_\mu(n, L, \mu, g_2)$$

Since this operation is based on the same congruence closure operation as the abstraction operation (plus a linear amount of work to compute the needed properties) the merge operation it can be computed in $O(E * log(N))$ time.

## 5 Additional Reduced and Interactive Views

While the abstract heap graph presented thus far produces models that scale in size with the number of logical regions in the program — independently of heap size and loosely correlated with the number of types used in the program — the graphs are often still too large to visualize and explore effectively. A second issue, particularly in a debugger scenario, is that after identifying a region of interest the developer wants to zoom into a more detailed view of the objects that make up the region.

While the DGML viewer [11] we use is quite effective at zooming, slicing, and navigating though large graphs we can directly address the above two issues by providing additional support for zooming between abstraction levels: the developer can zoom incrementally from a very high level view based on *dominators* in the abstract heap graph, defined in subsection 2.3, all the way down to individual objects in the concrete heap without losing track of the larger global context of the heap structure[3].

Given an abstract heap graph we can compute *dominator* information in a fairly standard way [24]. We deviate slightly since we want to ensure that *interesting* nodes which are directly pointed to by variables, and nodes that are immediate neighbors of these nodes remain expanded. In our experience this heuristic seems to strike a nice balance between collapsing large portions of the graph, to aid in quickly getting a general overview of the heap, while preserving structure around local variables, which are frequently of particular interest and we want extra detail on. This can be done by simply asserting that all of the nodes we want to keep expanded do not have any non-self dominators (equivalently ignoring all in-edges to these nodes during dominator computation). Using our modified dominator computation we can replace every node $n$ (which has not been marked *interesting*) and all of the nodes $n_1^d \ldots n_k^d$ that $n$ *dominates* with a single *reduced node*. This simple transformation results in a substantial reduction in the size of the graph while preserving much of the large scale heap structure and, since we can track the set of abstract graph nodes that each *reduced node* corresponds to, we can move easily between the two views of the heap. Furthermore, since the notion of *domination* and *ownership* [6] are closely related, this reduction has a natural relation with the developer's concept of ownership encapsulation of heap structures. This view is conceptually similar to the approach taken in [22,21], although the dominator construction is on the abstract graph, where data structures have already been identified and grouped, instead of on the concrete heap graph.

---

[2] Self-edges need not be considered as they are already accounted for in the shape.
[3] In a way that is similar to the *semantic zoom* of [8].

*Individual Object Zoom* When looking at a graph that represents an abstraction of a single heap state (e.g., in an interactive debugger) it is very useful to be able to zoom down from the level of individual regions to examine the individual objects that make up a region. One approach for this is to simply expand a node in the abstract graph into the concrete object graph it represents. However, for large structures (e.g., a list with 2000 entries) this can produce an intractably large graph. An alternative is to mark individual objects as *interesting* and then implement the abstraction function such that these objects are always represented as distinct nodes (i.e., never merged). Then as the user drills down into a data structure, similar to what is done in existing debuggers, we can recompute the abstraction for the data structure that is being explored marking the appropriate nodes as *interesting* so they can be individually inspected.

## 6  Implementation and Evaluation

To evaluate the utility of our abstraction, we examine 1) the cost of computing abstract heaps from realistically sized heaps in real programs, 2) the feasibility of visualizing the abstract graphs, and 3) whether the abstract graphs produced are precise enough for understanding the program's behavior and to identify and correct various defects.

We implemented the algorithms[4] for computing and manipulating abstract heap graphs in C#. In order to visualize the resulting graphs we use the DGML [11] graph format and the associated viewer in Visual Studio 2010. This graph format and viewer support conditional style macros to control changes between the levels of abstraction described in this paper, and to perform selective highlighting of nodes/edges with given properties. For example, we can highlight edges that represent *non-injective* pointers, or we can apply a *heat-color* map to the nodes based on the amount of memory the objects they represent are using.

---

[4] Code available online at `http://heapdbg.codeplex.com` and a web accessible demo is available at `http://rise4fun.com/HeapDbg`.

In order to evaluate the utility of the abstraction in the inspection and understanding of heap related problems (and in their solutions) we implemented a memory profiler tool. This profiler rewrites a given .Net assembly with sampling code to monitor memory use and to compute heap snapshots, along with the associated abstractions, at points where memory use is at a high point in the execution profile. The rewriter is based on the Common Compiler Infrastructure (CCI) [4] framework. As performing full heap abstractions at each method call would be impractical we use a per-method randomized approach with an exponential backoff based on the total reachable heap size (as reported by the GC). If we detect that the program may have entered a new phase of computation, the reachable heap size grows or shrinks by a factor of $1.5\times$ from the previous threshold, then we begin actively taking and abstracting heap snapshots. A snapshot of the heap is the portion reachable from the parameters of a single method call and from static roots. Depending on the size of the snapshot relative to previously seen heaps, we either save the snapshot as likely capturing some interesting heap state or discard it and increase the random backoff for the method that produced it. This use of random backoff sampling based on GC reported memory use and snapshot size results in a program that outputs between 2 and 10 snapshots from a program execution and execution is around $20\times$ to $100\times$ slower than the uninsturmented program. We compared the results obtained by sampling uniformly at random and found that, in addition to having a much larger overhead, the uniform sampling approach produced results that were no more useful for memory debugging then the backoff sampling approach.

In order to help the developer quickly identify structures of interest we implemented a number of simple post-processing operations on the abstract graphs which allow the DGML viewer to flag nodes (regions) of the heap that display common types of poor memory utilization [23]. The properties we identify are percentage of memory used, *small object* identification, *sparse container* or *small containers*, and *over-factored classes*. The memory percentage property uses a heat map, coloring any nodes that contain more than 5%, 15%, or 25% of the heap respectively. The small object property highlights any nodes where the object overheads (assumed to be 4 bytes per object) are more than half the actual data stored in the objects. The poor collection utilization property highlights nodes that represent regions which are containers and all of them are either all very small (contain 3 or fewer elements) or are more than half empty (over half the entries are null). While the first three properties are fairly standard, the final property, over-factored classes, is a less well known issue. We consider a structure overfactored if (1) there exists a node *n* that consists of small objects and (2) *n* has a single incoming edge that is *injective* (i.e., each object represented by the node *n* is uniquely owned by another object). These two features appear commonly when the objects represented by the node *n* could be merged with the objects that have the unique pointers to them (i.e., the class definitions can be merged) or when the objects represented by *n* could be better implemented as *value types* (i.e., `structs` in C#). The `Face[]` and `Point` objects in the raytracer study, subsection 6.1, are an example of this.

From the viewpoint of a userspace tool handling the types provided by the base class or system libraries, e.g., the Base Class Library (BCL) for .Net or the `java.*` in Java, are an important consideration. For user space applications the internal structure of say, `FileStream` or `StringBuilder` is not interesting, We identify these objects by simply examining the namespace of the type and treat them as single opaque objects. However, some classes in these libraries have features that are relevant to userspace code even though the details of the internal representation are not of particular interest. Examples of these types are `List<T>` or `Dictionary<K,V>`, which we treat as ideal algebraic data structures, showing the links to the contained elements but still treating the internal implementations as opaque.

For this paper we converted raytracer from from SPEC JVM98 [31] and six programs from DaCapo suite [3] to .Net bytecode using the ikvm compiler [13][5]. As the DaCapo suite contains a number of large and complex programs we also timed the extraction, comparision, and merge operations on each heap snapshot that was taken.

### 6.1 Raytracer: Extended Case Study

In this section we study the raytracer program from SPEC JVM98. The program implements a raytracer which renders a user provided scene. Using this example, we illustrate how the heap visualization looks for a well know program, and how the information can be used in a debugging type scenario to investigate memory use.

Running this program in the heap profiler, we obtain as one of the snapshots an abstract heap from the entry of the `shade` method. This abstract heap represents ∼168K objects (a total of ∼4MB of memory). Applying the heap graph abstraction followed by the dominator reduction produces the visualization shown in Figure 6.1. This figure shows the *entire* heap structure for the render while preserving most structural features of interest. In this heap we see the root nodes `this`, `tree`, and `eyeRay` representing the argument variables to the method. The `this` variable refers to a `scene` object. This object has a field `octree` that represents a space decomposition tree structure which is also referred to by the `tree` argument variable. The larger nodes with the *chevron* are *dominator reduced* nodes that represent multiple dominated regions and can be expanded to inspect the internal structure in more detail.

The raytracer octree space decomposition structure is represented by the dominator reduced node labeled `#20`. It is directly noticeable that there are pointers from this data structure to `ObjNode` objects, represented by node `#7`. The shape tree{nextLink} of node `#7` indicates that this is a list (a tree with out-degree 1). The list in turn contains shapes (`SphereObj`, `TriangleObj`, . . .) that are in the associated quadrants of the space decomposition structure. This list is used to enumerate all the shapes that appear in a given quadrant. There are also references from objects in the space decomposition tree structure to the dominator reduced node `#19`, which contains more information on the composite structure of `Face` objects.

*Memory Use.* Memory usage is an important concern for many applications. There are many reasons why an application may use more memory than is really required. Common problems in object-oriented, garbage collected languages are *leaks* [14], where unused objects are still reachable, and *bloat* [23], where encapsulation and layering have added excessive overhead. Ideally a programmer would like to see what types are using the most memory and where these objects are being used. Our visualization uses the conditional styling support in the DGML renderer to color nodes based on the percentage of total used live memory. Enabling this coloring results in the dominator reduced node representing the `Face` structures (node `#19`) being colored.

---

[5] Unfortunately, ikvm is not able to process the remaining DaCapo benchmarks.

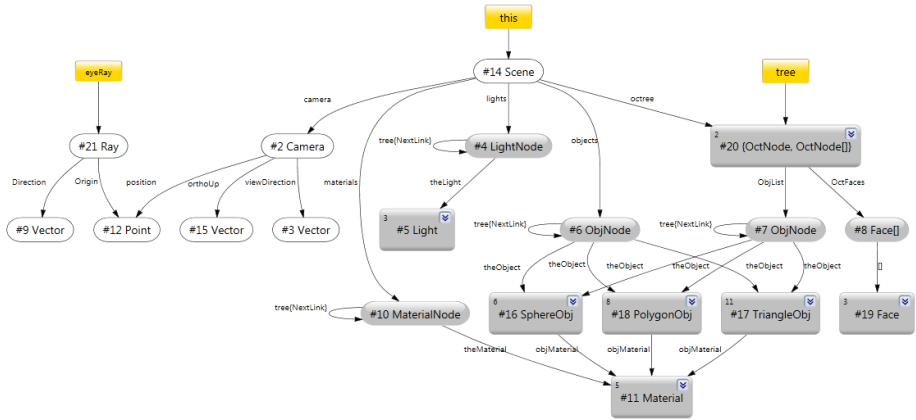

**Fig. 3.** Debugger snapshot of `Shade` method in the `Scene` class.

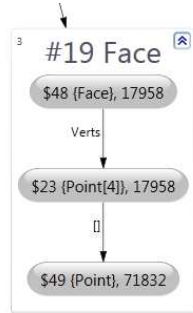

**Fig. 4.** Expanded face dominator node.

Node `#19` represents a large amount of memory, ∼107K objects representing nearly half of the the total live heap. By expanding the node `#19` we get the graph (Figure 4) representing the internal structure of the dominator reduced node. This reveals node ($48), abstracting a region of ∼18K `Face` objects, node ($23), abstracting a region of ∼18K `Point[]`, and node ($49), abstracting a region of ∼72K `Point` objects. The raytracer program is known to have poor *memory health* [23], in the sense that it exhibits a high rate of object overhead associated with a large number of very small objects. The `Point` objects here are a major factor in that.

At first glance it may not be clear how to reduce the overhead of these `Point` objects. However, turning on the *over-factored* highlighting or inspecting the *injectivity* information in Figure 4, provides additional guidance. The edge from node $23 to node $49—representing all the pointers stored in the arrays—is shown as a normal edge and not shaded and wide. Therefore, the set of pointers abstracted by the edge is *injective* and each index of each array points to a unique `Point` object. Given this likely ownership relation and the fact that all of the arrays are of length 4 it seems that flattening the *Face* data structure would reduce memory use substantially (i.e., this satisfies our conditions for being an *over factored* structure).

By studying the source code for the `Face` class we can see that these ownership and length properties do in fact hold universally. Thus, we can flatten each `Point[4]` and associated `Point` objects into a `float[12]`. This transformation eliminates one object header per `Point` object (at 4 bytes each) and the 4 pointers stored in the `Point[4]` (at 4 bytes per pointer). Given that we have ∼72K `Point` objects and ∼18K `Point[]`, this change works out to ∼0.6MB of savings or ∼18% of the total live heap. Using similar reasoning we could further flatten the `float[12]` arrays into the `Face` implementations for another ∼0.22MB of savings, or another ∼6% of the live heap. These two refactorings then represent a 24% reduction in the total live heap size.

This case study shows how the multi-level abstraction allows the developer to navigate around the heap at the desired level of detail, zoom-in and out of specific areas of interest, all while maintaining the larger context. This ability to put the problem in context and interactively explore the heap is critical to aiding the developer in quickly identifying the source of a problem, understanding the larger context, and thus being confident in formulating a remedy.

### 6.2 Evaluation With Profiler

A number of papers have identified and explored memory use issues in the DaCapo benchmark suite. Hence, we decided to evaluate the effectiveness of the abstraction techniques described in this paper by using our profiling tool to analyze several programs from the DaCapo suite for memory utilization issues.

After running the profiler we inspected the output abstract graphs to find nodes (regions) that contained potential problems and then to determine what (if anything) could be done to resolve the issues or if the memory use appeared appropriate. This was done via manual inspection of the graph, the use of the heap inspection and highlighting tools in the profiler, and inspecting the associated source code. In all cases at most 7 nodes were colored by the profiler tools and the total time to inspect the graph, identify the relevant structures, inspect the associated source code, and determine if the memory use was generally appropriate was always less than 10 minutes. Also, as we had not previously worked with the code, sometimes we needed to spend additional time to understand more about the intent of the classes and their structure in order to fully determine if the code could be successfully refactored and how. This was particularly important when multiple classes/subclasses were used to build recursive data structures. However, this inspection never required more than an additional 15 to 20 minutes.

*Antlr.* For the `Antlr` benchmark, the tool reports one of the larger heaps being reachable from a method in the `JavaCodeGenerator` class. We inspected this heap with our visualization turning on the memory use heat map, we were able to quickly identify one dominator node as containing around 72% of the reachable memory. This region was dominated by a set of `RuleSymbols` each of which stores information representing various aspects of the parser. Further inspection did not reveal any obvious memory use problems or obvious areas where data structures could be refactored to substantially improve memory utilization. These findings match those of previous studies of the benchmark which is not known to have any reported memory leaks and is reported to have good utilization of memory (in particular [23] reports a good *health* score).

*Chart.* For the `Chart` benchmark our tool reports the largest heaps being reachable from a method in the `JFreeChart` class. Our highlighting tools indicate a region, Figure 5, that is of potential interest is dominated by a set of `XYSeries` objects. Expanding this dominator node shows that the memory is being used by a large number of `XYDataItem` objects and the `Double` objects they own (similar to the case in the raytracer case study). By hovering over these objects we saw that they consume about 3MB of heap space. The actual data contained in these objects (in particular the `Double` objects) is small compared to the object overhead and there is an ownership relation between each of the `XYDataItem` objects and the Double objects. This indicates that we could inline these structures to save space and an inspection of the `XYDataItem` class shows that it declares the `x/y` fields as `Number` types to allow for some level of polymorphism. So we need to subclass our new flattened classes to allow for storing both integer and floating point x,y pairs. This refactoring results in a savings of around 1MB, which is around 25% of the total live memory at this point in the program. To the best of our knowledge this memory issue has not been reported in previous work.

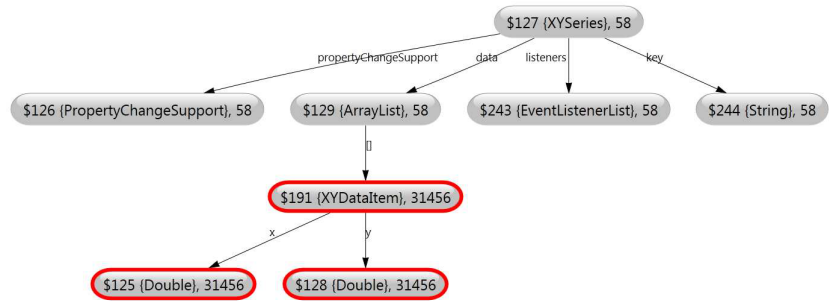

**Fig. 5.** Chart Memory Use

*FOP.* For the `fop` benchmark the tool reports the largest heap being reachable from a method in the `Page` class. The highlighted region consists of a large number of objects that contains various parts of the document to render, for example, the `WordArea` and `TableArea` objects. After a some inspection of the source code we concluded that the data structure was not particularly amenable to refactoring. As reported in [14], we note that the data structure is needed later in the computation and thus is not a leak.

*PMD.* For the `pmd` program, our tool reports one of the larger heaps as occurring in the `JavaParser` class. The section highlighted by the memory utilization coloring uses over 10MB of memory and consists of a data structure which is a tree via the `children` field and container offsets, along with a parent back pointer on the `parent` field. This data structure represents the AST of the program that is being analyzed. Hovering over the node reports that it represents more than 50 types (with names like `ASTExpression` and `ASTPrimitiveType`) that all inherit from the `SimpleNode` class. On inspection we see that this base class has many data fields (line numbers, the `children` array, the `parent` field, etc.) which the subclasses add to with additional AST specific information. Given this structure we did not see any obviously poor memory use or memory leaks. This appears to contradict [23] which reports a high rate of object header overhead in this benchmark. However, in this case the overhead is actually encoding important structural information about the AST that is being processed. This demonstrates how the visualization can be used to augment the information provided by existing analysis tools.

### 6.3 Computational Costs

Table 1 contains information on the sizes of the largest abstract representations produced during the runs of the profiler and the cost of extracting and comparing these abstract heap graphs. The first column lists the benchmark and the second column the number of objects in the largest concrete heap snapshot that was encountered. The following columns are the size of the largest abstract heap graph produced for any heap snapshot (*AbsNode*), and the size of the corresponding dominator reduced representation from section 5 (*Reduced*). Some of these sizes seem to be at (or beyond) the upper end of what can be conveniently visualized. However, our experience with in subsection 6.1 shows the combination of the conditional graph styles, the ability to zoom between levels of detail, and the navigational tools provided by the DGML viewer made inspecting and understanding the relevant parts of the graphs quite easy.

| Bench | Objects | AbsNode | Reduced | AbsTime | EqTime | MergeTime |
|---|---|---|---|---|---|---|
| raytracer | ~168K | 48 | 21 | 1.37s | 0.04s | 0.11s |
| antlr | ~12K | 606 | 201 | 0.41s | 0.03s | 0.11s |
| chart | ~189K | 198 | 110 | 3.22s | 0.09s | 0.21s |
| fop | ~120K | 531 | 150 | 2.67s | 0.11s | 0.41s |
| luindex | ~2K | 87 | 36 | 0.50s | 0.01s | 0.02s |
| pmd | ~178K | 146 | 28 | 4.11s | 0.09s | 0.15s |
| xalan | ~40K | 451 | 127 | 2.42s | 0.07s | 0.17s |

**Table 1.** Max graph sizes and timings.

The next issue we wanted to evaluate was the computational costs of performing the abstraction, comparison, and merge operations. The columns *AbsTime*, *EqTime*, and *MergeTime* columns shows the maximum time taken to abstract a concrete heap during the profiler run and to merge/compare it with previously taken snapshots.

The current abstraction implementation creates a complete shadow copy of the concrete heap during abstraction. Despite this large constant time overhead, the cost of computing the abstractions is quite tractable. The running time scales very closely to the asymptotic complexity of $O(E * log(N))$. The current implementation computes the abstraction inside the process that is instrumented, so it was not possible to precisely measure the exact memory overhead of the abstraction operations. However, using the difference in the total memory consumed by the process as reported by the system monitor indicates a factor of a 40× increase in memory use (never exceeding 800MB).

## 7 Related Work

Developing debugger support for the program heap is an ongoing active research area. The work in [33] outlines many of the basic issues that arise when attempting to visualize concrete program heaps and [27] presents some abstractions to help alleviate some of these issues. There is a large body of work on techniques to improve the efficiency and effectiveness of debugging [32,17,18,29,12,26]. Work in [1] takes the same general approach as this work but focuses on the interactive aspects of visualizing the heap, in particular allowing the developer to inspect individual objects in a larger structure.

Work by Mitchell et. al. [22,21] has a number of similarities to the work in this paper. Both approaches use a set of grouping heuristics to partition structures in the heap and then extract information about the partitions, but the partitioning strategy and information extracted differ substantially. Our work uses recursive structures and predecessor ownership to identify equivalence classes of objects/data while [22,21] focus on dominator relations between objects. We note that this results in the same asymptotic cost as the work in this paper. Given this difference of grouping heuristics there is also a natural difference in the focus on what type of information is extracted. In particular, the abstraction in this paper is designed to aid programmer understanding of the structure and connectivity of various heap structures and so it explicitly extracts information on shape, edge injectivity, pointer nullity, container sizes, in addition to information on the sizes of various data structures. While some of these properties can, in some cases, be reconstructed using *fanout* and object count information, the majority of the information computed in [22,21] focuses the specific task of identifying memory inefficiencies in large Java programs.

There is a substantial amount of work on the development of heap models for use in static program analysis [2,10,16,30]. Whereas program analysis is concerned with computability and obtaining enough precision at reasonable cost, the main challenge in abstracting runtime heaps is to obtain very small models that can be visualized, while retaining many useful properties of the original heap. We believe though that insights in static heap analysis can inform the abstractions of runtime heaps and vice versa. For example, it would be interesting to provide programmers with more control over the abstractions produced via instrumentation predicates [2,30]. The approach in [16] uses a less descriptive model than the one presented in this paper for example, it does not consider information such as injectivity or shape. Work in [28,15] use a related idea of taking a concrete heap from a C/C++ or Java program and inferring the types [28] or basic shapes [15] of heap structures.

## 8 Conclusion

This paper introduces a new runtime technique for program understanding, analysis and debugging. The abstraction of heap graphs presented attempts to construct a very small representation of the runtime heap in order to allow effective visualization and navigation, while retaining crucial high-level properties of the abstracted heap, such as edge relations and shape of various subgraphs. The construction of the abstraction ensures that the abstract graph is a *safe* representation of the concrete heap, allowing the programmer (or other tools) to confidently reason about the state of the memory by looking at the abstract representation. Our benchmarks and case studies demonstrate that abstract heap graphs can be efficiently computed, contain interesting information on the heap structure, and provide valuable information for identifying and correcting memory use related defects. Given the utility of the abstraction in this task we believe there are a number of other applications including thread races, refactoring for parallelism, or interactive debugging, where this type of abstraction and understanding would be useful.

## Acknowledgments

We would like to thank Peli de Halleux for setting up the online demo code at RiSE4fun, Chris Lovett for his help with DGML, and Todd Mytkowicz for his insight into some of the code in the DaCapo suite.